\documentclass[10pt,conference]{IEEEtran}
\IEEEoverridecommandlockouts

\usepackage{cite}
\usepackage{amsmath,amssymb,amsfonts}
\usepackage{algorithmic}
\usepackage{graphicx}
\usepackage{textcomp}
\usepackage{xcolor}
\usepackage{tikz, pgfplots}
\pgfplotsset{compat=1.17}
\usepgfplotslibrary{groupplots}
\usepackage{siunitx}
\usepackage{url}
\usepackage{bm}
\def\BibTeX{{\rm B\kern-.05em{\sc i\kern-.025em b}\kern-.08em
    T\kern-.1667em\lower.7ex\hbox{E}\kern-.125emX}}

\renewcommand{\vec}[1]{\mathbf{\bm{#1}}}
\begin{document}

\title{Sound Field Reconstruction Using Physics-Informed Boundary Integral Networks\\

\thanks{This project has received funding from KU Leuven internal funds C3/23/056 and from FWO Research Project G0A0424N.}
}
\author{
    \IEEEauthorblockN{Stefano Damiano, Toon van Waterschoot}

    \IEEEauthorblockA{Dept. of Electrical Engineering (ESAT-STADIUS), KU Leuven, Leuven, Belgium
    \\\{stefano.damiano, toon.vanwaterschoot\}@esat.kuleuven.be}
    
}

\maketitle

\begin{abstract}
Sound field reconstruction refers to the problem of estimating the acoustic pressure field over an arbitrary region of space, using only a limited set of measurements. Physics-informed neural networks have been adopted to solve the problem by incorporating in the training loss function the governing partial differential equation, either the Helmholtz or the wave equation. In this work, we introduce a boundary integral network for sound field reconstruction. Relying on the Kirchhoff-Helmholtz boundary integral equation to model the sound field in a given region of space, we employ a shallow neural network to retrieve the pressure distribution on the boundary of the considered domain, enabling to accurately retrieve the acoustic pressure inside of it. Assuming the positions of measurement microphones are known, we train the model by minimizing the mean squared error between the estimated and measured pressure at those locations. Experimental results indicate that the proposed model outperforms existing physics-informed data-driven techniques.
\end{abstract}


\section{Introduction}
\label{sec:introduction}
Immersive audio technologies aiming to create a realistic auditory experience in virtual and augmented reality demand a precise spatial representation of acoustic environments, in order to enhance the user engagement~\cite{zotterAmbisonics2019}. Obtaining direct measurements in a large space is a demanding process~\cite{dietzenMYRiADMultiarrayRoom2023}, that can rarely be pursued in practice. As a solution, sound field reconstruction aims at estimating the acoustic pressure in areas where direct measurements are not available, given a limited set of microphone recordings. 

The problem has been addressed using model-based and data-driven methods. Model-based solutions either rely on a parametric representation of the acoustic scene~\cite{pezzoliParametricApproachVirtual2020} or describe the sound field using closed-form solutions to the wave equation~\cite{UenoKernelRidgeRegression2018}, including plane waves~\cite{jin2015theory}, spherical waves~\cite{borraSoundfieldReconstructionReverberant2019}, and equivalent sources~\cite{antonelloRoomImpulseResponse2017,koyamaSparseRepresentationSpatial2019,damianoCompressiveSensingApproach2024}. These models, while robust and interpretable, are based on assumptions and specific propagation models, that hinder their capacity to capture the complexity of the sound field over a considerably large area inside a room. Data-driven techniques have thus been explored to learn representations of the sound field from extensive datasets. Deep neural networks have proven effective in learning implicit representations of the acoustic field~\cite{suINRASImplicitNeural2022,luoLearningNeuralAcoustic2022}, achieving good performance in sound field reconstruction~\cite{fernandez-grandeGenerative2023,miotelloReconstructionSoundField2024}. Moreover, dictionary learning techniques have been investigated to represent the audio signal as a combination of functions (called dictionary atoms) learned from data~\cite{hahmannSpatialReconstructionSound2021}. As a drawback, both types of data-driven methods necessitate substantial data for training and lack interpretability.

To bridge the gap between model-based and data-driven methods, physics-informed machine learning approaches have been proposed, exploiting prior knowledge of the physical laws that govern sound propagation in the learning process~\cite{koyama2024physics}. 
Physics-informed neural networks (PINNs) integrate in the training objective the underlying partial differential equation (PDE), either the wave equation, in the time domain~\cite{pezzoliImplicitNeuralRepresentation2024,sundstrom2024sound,olivieri2024physics,karakonstantisRoomImpulseResponse2024}, or the Helmholtz equation, in the frequency domain~\cite{shigemiPhysicsInformed2022,ribeiro2024sound,maSoundFieldReconstruction2024}.
This is achieved by introducing in the loss function a term that penalizes the deviation of the estimated sound field from the underlying PDE, evaluated at a dense set of collocation points in the target area. As a result, these networks are trained using solely a sparse set of observations. Similarly, in~\cite{damianoZeroShot2024} a zero-shot physics-informed dictionary learning technique, not requiring pre-training data, has been introduced.

Recently, physics-informed boundary integral networks (PIBI-Nets) have been proposed to solve PDEs~\cite{linBINetLearningSolve2021, linBIGreenNetLearningGreens2023,sunBINNDeepLearning2023,nagy-huberPhysicsinformedBoundaryIntegral2024}. These models are derived from the boundary element method (BEM)~\cite{kirkupBoundaryElementMethod2019}: rather than directly exploiting PDEs, in fact, the associated \textit{boundary integral equation} (BIE) is used to guide the training. It has been shown that PIBI-Nets outperform PINNs in the solution of Laplace and Poisson problems~\cite{nagy-huberPhysicsinformedBoundaryIntegral2024}.

In this work, we investigate PIBI-Nets for sound field reconstruction. According to the Green's theorem, the sound field in a closed area can be expressed, using the Kirchhoff-Helmholtz BIE, as a function of the pressure density on its boundary~\cite{williamsFourierAcousticsSound1999,kirkupBoundaryElementMethod2019}. 
While the BEM relies on the solution of a linear system, the proposed method,
inspired by~\cite{nagy-huberPhysicsinformedBoundaryIntegral2024}, employs a multi-layer perceptron (MLP) to learn the pressure density at a discrete set of integration points on the boundary of the reconstruction region. The boundary density is then used to estimate, by means of the Kirchhoff-Helmholtz BIE, the sound field inside the region.

While PINNs require a dense set of collocation points over the reconstruction region, the proposed solution relies on integration points only on the boundary: the dimensionality and complexity of the problem are thus reduced, facilitating training. Experimental results indicate that the proposed method outperforms both the physics-informed dictionary learning approach~\cite{damianoZeroShot2024} and a traditional PINN, is more robust to a decreasing number of available measurements, and requires fewer integration points compared to the collocation points required by the PINN. Finally, the proposed method does not entail prior knowledge on the source position, room geometry or boundary conditions, and produces an interpretable solution.

\section{Problem Formulation and Background}
\label{sec:problem_formulation}
\subsection{Problem formulation}
\label{subsec:problem_formulation}
Let us consider a reverberant room containing one or more acoustic sources, and a two-dimensional, source-free region of space $\Omega \in \mathbb{R}^2$ lying on a horizontal plane inside the room. We can express by $p(\vec{r}, \omega) \in \mathbb{C}$ the complex-valued acoustic pressure in the frequency domain, at position $\vec{r} \in \Omega$ and angular frequency $\omega = 2\pi f$, where $f$ is the temporal frequency. The frequency-domain sound field satisfies the homogeneous Helmholtz equation~\cite{williamsFourierAcousticsSound1999}
\begin{equation}\label{eq:helmholtz_pde}
    (\nabla^2 + k^2)p(\vec{r},\omega) = 0\,,
\end{equation}
where $\nabla^2$ denotes the Laplace operator and $k=\omega/c$, with $c$ being the speed of sound in air. We sample the acoustic pressure in $\Omega$ using $M$ microphones, whose positions are indicated by $\{\vec{r}_m\}_{m=1}^M$. The sound field measured at the $m$th microphone can be expressed as 
\begin{equation}
    s(\vec{r}_m, \omega) = p(\vec{r}_m, \omega) + e_m\,,
\end{equation}
where $e_m$ denotes additive measurement noise. We can collect the $M$ microphone measurements in the vector $\vec{s} \in \mathbb{C}^M$,
\begin{equation}
    \vec{s} = [s(\vec{r}_1, \omega), s(\vec{r}_2,\omega),\ldots, s(\vec{r}_M, \omega)]^\top\,.
\end{equation}
Our goal is to retrieve the pressure field at arbitrary positions within $\Omega$, given the available measurements $\vec{s}$. For the sake of readability, we omit the dependency on the frequency $\omega$ in the rest of the paper.

\subsection{Background}
\label{subsec:background}
Traditional data-driven methods for sound field reconstruction rely on learning from data a function $f(\vec{r}, \vec{\theta})$, parametrized by the vector $\vec{\theta}$, that maps spatial coordinates in $\Omega$ to the corresponding pressure value. This function can be parametrized using a neural network trained to minimize a data-fidelity loss function, expressing how much the estimated pressure at the positions of the available microphones $\{\vec{r}_m\}_{m=1}^M$ deviates from the actual measurements $\vec{s}$. The problem can be framed as
\begin{equation}
    \vec{\theta}^* = \arg\min_\vec{\theta} \mathcal{L}_\text{data}\left(f(\{\vec{r}_m\}_{m=1}^M, \vec{\theta}), \vec{s}\right)\,,
\end{equation}
where the loss function $\mathcal{L}_\text{data}$ usually takes the form of a mean squared error
\begin{equation}
    \mathcal{L}_\text{data}\left(f(\{\vec{r}_m\}_{m=1}^M, \vec{\theta}), \vec{s}\right) = \sum_{m=1}^M\left\vert s(\vec{r}_m) - f(\vec{r}_m, \vec{\theta}) \right\vert^2\,.
\end{equation}
To enable $f$ to effectively generalize to positions where microphones are not available (i.e., to prevent overfitting), regularization techniques are adopted in the literature. To encourage $f$ to produce a physically meaningful sound field, PINNs introduce an additional term in the loss function, penalizing solutions that deviate from the underlying PDE. This PDE loss term can be obtained from~\eqref{eq:helmholtz_pde} as 
\begin{equation}
    \mathcal{L}_\text{PDE} = \sum_{n=1}^N \left\vert (\nabla^2 + k^2)f(\vec{r}_n, \vec{\theta})\right\vert^2\,,
\end{equation}
where the PDE is evaluated at a set of $N$ points, usually called collocation points, that can be arbitrarily picked within $\Omega$, independently from the $M$ sampling points. The loss function used to train the neural network becomes
\begin{equation}
    \mathcal{L}_\text{PINN} = \mathcal{L}_\text{data} + \lambda \mathcal{L}_\text{PDE}\,,
    \label{eq:loss_pinn}
\end{equation}
where the positive constant $\lambda$ represents a weighting factor between the two terms, constituting a hyperparameter of the model. The advantage of PINNs is that the measurements $\vec{s}$ constitute the only data used to train the network, significantly reducing the need for labeled data, that represents a major drawback of traditional data-driven techniques.

\section{Proposed Method}
\label{sec:proposed_method}
Boundary integral networks aim at solving the PDE exploiting the boundary integral equation. Let us denote the boundary of $\Omega$ as $\partial\Omega$ and define the boundary density function $h(\vec{y})$, representing the pressure at some points $\vec{y}$ lying on $\partial\Omega$. An explicit integral representation of the solution of the homogeneous Helmholtz equation can then be written as~\cite{williamsFourierAcousticsSound1999} 
\begin{equation}
    p(\vec{r}) = \int_{\partial\Omega} \left[ G(\vec{y}, \vec{r})\frac{\partial h}{\partial \vec{n}_\vec{y}}(\vec{y}) - \frac{\partial G}{\partial \vec{n}_\vec{y}}(\vec{y}, \vec{r}) h(\vec{y})\right]\,d\sigma_\vec{y}\,,
    \label{eq:helmholtz_bie}
\end{equation}
where $\vec{n}_\vec{y}$ denotes the normal to $\partial\Omega$ at point $\vec{y}$ and $G(\vec{y},\vec{r})$ is the fundamental solution of the Helmholtz equation. This solution is represented by the Green's function, that defines the transfer function between a source located at position $\vec{y}$ and a receiver at position $\vec{r}$, and takes the form\cite{williamsFourierAcousticsSound1999}
\begin{equation}
    G(\vec{y},\vec{r}) = \begin{cases} 
        -\frac{j}{4}H_0^{(1)}(k\Vert\vec{r}-\vec{y}\Vert_2) & \text{2D case}\\
        \frac{1}{4\pi\Vert\vec{r}-\vec{y}\Vert_2}e^{jk\Vert\vec{r}-\vec{y}\Vert_2} & \text{3D case}\,.
    \end{cases}
\end{equation}
Here, $j$ denotes the imaginary unit, $\Vert \cdot \Vert_2$ is the $\ell_2$ norm and $H_0^{(1)}$ is the zeroth-order Hankel function of the first kind. Note that \eqref{eq:helmholtz_bie} relates the sound field at arbitrary points \textit{inside} $\Omega$ to the pressure at boundary points on $\partial\Omega$. Given that Green's functions and their derivatives are available in closed-form, and that $\partial\Omega$ is known (in fact, $\Omega$ is arbitrarily defined), inspired by~\cite{nagy-huberPhysicsinformedBoundaryIntegral2024} we adopt a neural network to estimate the unknown boundary pressure distribution $h(\vec{y})$. The network follows the same architecture presented in~\cite{nagy-huberPhysicsinformedBoundaryIntegral2024}, with 3 fully-connected layers of 64 neurons each and a single-neuron output layer. The network takes as input the spatial coordinate $\vec{r}$ of a position within $\Omega$ and outputs the value $h(\vec{y})$, then used to compute the estimated $\hat{p}(\vec{r})$ using~\eqref{eq:helmholtz_bie}. As the pressure is complex-valued, its real and imaginary parts are treated separately and then combined.
To compute the integral, a set of $N_\text{int}$ integration points are selected on $\partial\Omega$ and \eqref{eq:helmholtz_bie} is discretized similarly to~\cite{nagy-huberPhysicsinformedBoundaryIntegral2024} to compute the pressure within $\Omega$. To train the neural network, the sound field at the positions $\{\vec{r}_m\}_{m=1}^M$ of the $M$ microphones is estimated using PIBI-Net, and the deviation from the measured pressure is minimized using a mean squared loss function of the form
\begin{equation}
    \mathcal{L}_\text{PIBI} = \sum_{m=1}^M \left\vert s(\vec{r}_m) -\hat{p}(\vec{r}_m)\right\vert^2\,,
    \label{eq:loss_pibi}
\end{equation}
where $\hat{p}(\vec{r}_m)$ indicates the pressure estimated using PIBI-Net as in~\eqref{eq:helmholtz_bie}. Note that, differently from the PINN loss~\eqref{eq:loss_pinn}, $\mathcal{L}_\text{PIBI}$ has a unique term and does not contain any hyperparameter: thus, PIBI-Net has one hyperparameter less than PINN.

The boundary integral network comes with two main advantages over PINNs. First, as the predicted sound field is computed using~\eqref{eq:helmholtz_bie}, it always satisfies the Helmholtz equation: thus, the proposed model is interpretable and produces physically meaningful solutions. Note also that the neural network output $h(\vec{y})$ is constrained to be in the solution space of~\eqref{eq:helmholtz_bie} which, as discussed in~\cite{linBINetLearningSolve2021}, reduces the complexity of the problem and leads to an improved convergence. Second, while PINNs necessitate collocation points across the entire domain $\Omega$, PIBI-Nets require integration points only on its boundary $\partial\Omega$, which reduces by 1 the dimensionality of the problem. Consequently, as will be shown in Sec.~\ref{sec:evaluation}, a PIBI-Net achieves better performance with a lower $M$ and with fewer integration points than the collocation points needed by a PINN.

\section{Evaluation}
\label{sec:evaluation}
\begin{figure}
    \centering
    \includegraphics[width=0.88\columnwidth]{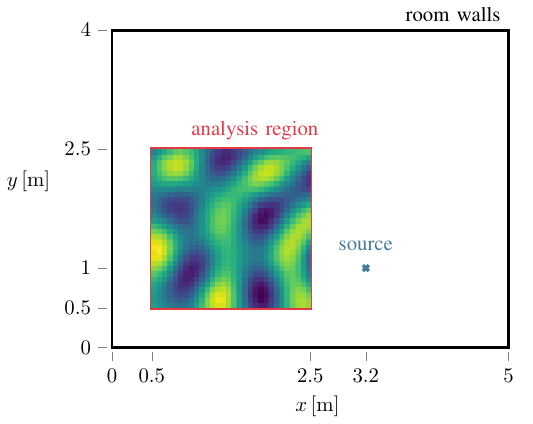}
    \caption{Geometry of the simulated room considered in the experiments.}
    \label{fig:room_geometry}
\end{figure}
To assess the performance of the proposed method, we implement it in Python\footnote{Code available: \texttt{\url{https://github.com/steDamiano/pibi-sfr}}} and run an experimental campaign on simulated data. Using pyroomacoustics~\cite{scheiblerPyroomacousticsPythonPackage2018}, we create a 2D rectangular room with size $[5,4]\,\si{\meter}$, having $T_{60}=\SI{0.4}{\second}$ and containing a single, omnidirectional source located at position $[3.2, 1]\,\si{\meter}$. We choose our analysis region $\Omega$ to be a square with $\SI{2}{\meter}$ side, with its bottom right corner at position $[0.5, 0.5]\,\si{\meter}$. In this area, we define a grid of $30\times30$ equally spaced points, simulate room impulse responses (RIRs) between the source and these points, with a sampling frequency $f_s=\SI{16}{\kilo\hertz}$, and transform them to the frequency domain using the fast Fourier transform (FFT) with $1024$ frequency bins. The simulated scenario is illustrated in Fig.~\ref{fig:room_geometry}. Note that this 2D geometry corresponds to a shoebox room having totally absorbing ceiling and floor, with the source and all microphones located on the same horizontal plane.

To evaluate the proposed method, we use two different baselines: the physics-informed dictionary learning method~\cite{damianoZeroShot2024} (PIDL) and a physics-informed neural network (PINN) having the same architecture of the proposed method (PIBI), described in Sec.~\ref{sec:proposed_method}. The PIDL can be directly applied to the grid defined in the reconstruction area, and its hyperparameters are set as in~\cite{damianoZeroShot2024}. We implement the PINN and PIBI networks in Pytorch~\cite{paszkePytorchImperativeStyle2019} and train them for $5\,000$ steps using the Adam optimizer, a learning rate of $0.001$, and the loss functions defined in~\eqref{eq:loss_pinn}, with manually tuned $\lambda=0.001$, and~\eqref{eq:loss_pibi}, respectively. Both architectures have $8\,577$ parameters.

As we consider a source-free reconstruction region, we adopt the homogeneous Helmholtz PDE~\eqref{eq:helmholtz_pde} for PINN, and its boundary integral representation~\eqref{eq:helmholtz_bie} for PIBI. When training the PINN, the partial derivatives of the PDE are computed using automatic differentiation, available in Pytorch.
\begin{figure*}[t]
    \centering
    \includegraphics[width=\linewidth]{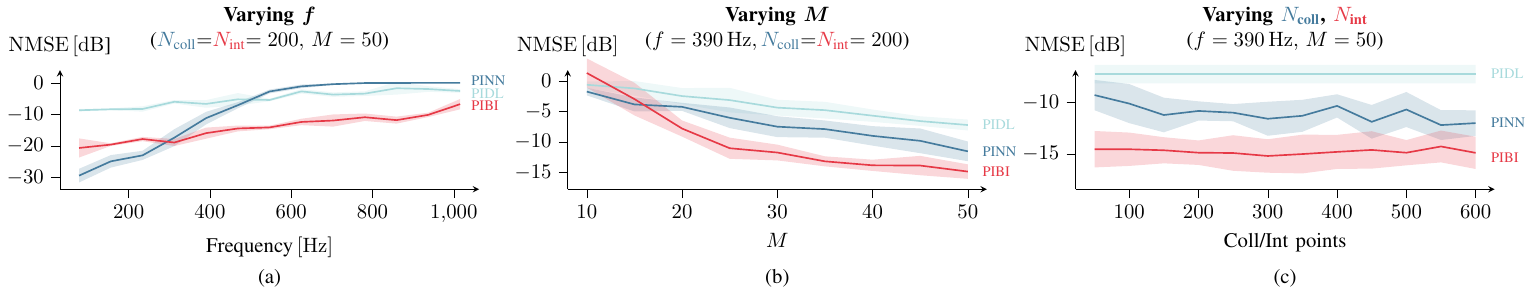}
    \caption{Mean (solid line) and standard deviation (shaded region) of the $\operatorname{NMSE}$ scores, obtained with the baselines PIDL and PINN and the proposed method PIBI when varying: (a) the frequency; (b) the number of measurement microphones $M$; (c) the number of collocation (PINN) or integration (PIBI) points.}
    \label{fig:nmse_plot}
\end{figure*}
\begin{figure*}[t]
    \centering
    \includegraphics[width=\linewidth]{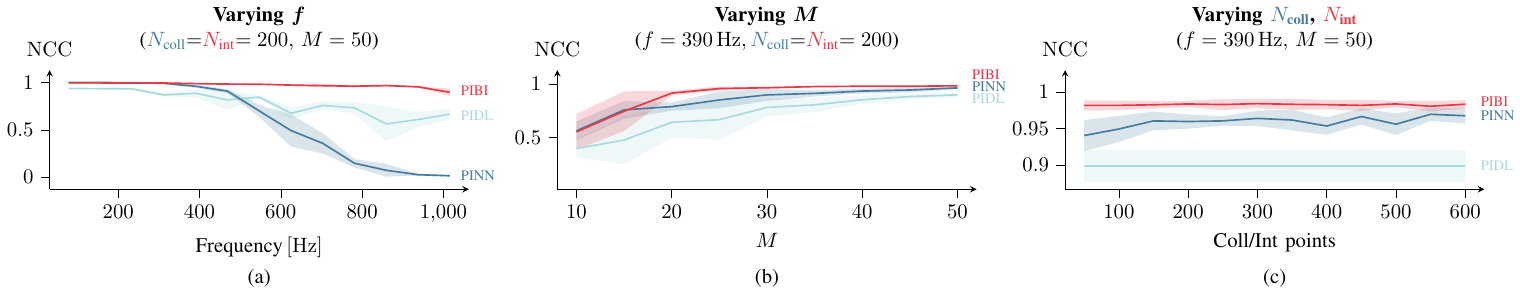}
    \caption{Mean (solid line) and standard deviation (shaded region) of the $\operatorname{NCC}$ scores, obtained with the baselines PIDL and PINN and the proposed method PIBI when varying: (a) the frequency; (b) the number of measurement microphones $M$; (c) the number of collocation (PINN) or integration (PIBI) points.}
    \label{fig:ncc_plot}
\end{figure*}
\begin{figure*}[t]
    \centering
    \includegraphics[width=0.97\linewidth]{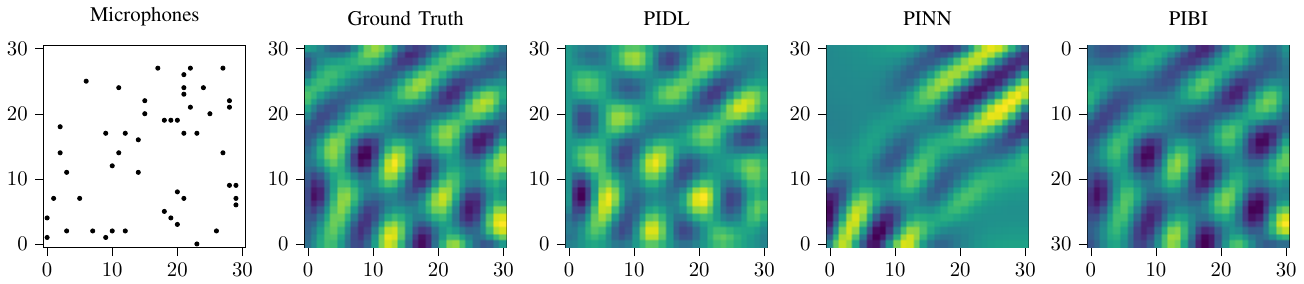}
    \caption{Real part of the sound field at $f=\SI{625}{\hertz}$ on the $30\times30$ point grid, obtained with the PIDL and PINN baselines and the proposed method PIBI, and compared to the ground truth. The problem is solved using $M=50$ measurements with microphones randomly arranged as in the leftmost figure.}
    \label{fig:sound_field_plot}
\end{figure*}

We adopt two different evaluation metrics: the normalized mean squared error (NMSE) in $\si{\decibel}$ scale
\begin{equation}
    \operatorname{NMSE} = 10\log_{10} \left( \frac{\lVert \vec{p} - \hat{\vec{p}} \rVert_2^2}{\lVert \vec{p} \rVert_2^2} \right) \,,
\end{equation}
where $\vec{p}$ contains the measured pressure and $\vec{\hat{p}}$ the predicted pressure at the $30\times 30$ grid points, and the normalized cross-correlation (NCC)
\begin{equation}\label{eq:ncc}
    \operatorname{NCC} = \frac{\lvert \hat{\vec{p}}^\mathrm{H}\vec{p}\rvert}{\lVert \hat{\vec{p}}\rVert_2^2 \lVert \vec{p}\rVert_2^2} \,,
\end{equation}
where $(\cdot)^\mathrm{H}$ indicates the Hermitian transpose. The $\operatorname{NCC}$ measures the similarity between the predicted and ground truth pressure, expressed by a value in the range $[-1, 1]$, where \(1\), \(0\), and \(-1\) correspond to perfect similarity, no correlation, and perfect anti-correlation, respectively \cite{damianoZeroShot2024}.

We compare the performance of PIDL, PINN and PIBI for varying frequency, number of microphones $M$, and number of collocation/integration points. We perform all experiments using a 5-fold procedure, where we randomly select different measurement microphones in each fold, and report the average and standard deviation of the results. In all experiments, the real and imaginary parts are reconstructed separately and combined to obtain the total reconstructed pressure. Following~\cite{nagy-huberPhysicsinformedBoundaryIntegral2024}, we first choose $N_\text{coll}=200$ collocation points in the reconstruction region for PINN, randomly selected at each iteration, and $N_\text{int}=200$ equally spaced integration points on the boundary for PIBI. We randomly pick $M=50$ microphones from the defined grid and analyze the reconstruction performance across frequencies ranging from $\SI{50}{\hertz}$ to $\SI{1050}{\hertz}$. In Fig.~\ref{fig:nmse_plot}a and Fig.~\ref{fig:ncc_plot}a we report, respectively, the $\operatorname{NMSE}$ and $\operatorname{NCC}$ scores for the three models. All models show a degradation of the performance as frequency increases: PIBI consistently outperforms PIDL at all frequencies on both metrics, and significantly outperforms PINN for frequencies higher than $\SI{350}{\Hz}$. In particular, the $\operatorname{NCC}$ score of PIBI is close to 1 (i.e., perfect similarity) across the entire range, whereas for PINN it gets close to zero (i.e., no correlation) for frequencies higher than $\SI{800}{\hertz}$. In Fig.~\ref{fig:sound_field_plot}, we show the real part of the sound field at $\SI{625}{\Hz}$, reconstructed in $\Omega$ by the three models and compared with the ground truth for $M=50$.

We then fix the frequency to $\SI{390}{\hertz}$ and set $N_\text{coll}=N_\text{int}=200$, and progressively reduce the number of measurement microphones; we report the performance of the three models in Fig.~\ref{fig:nmse_plot}b and Fig.~\ref{fig:ncc_plot}b. All methods exhibit a degradation of the performance as the number of available measurements decreases. Nonetheless, PIBI is more robust than the two baselines to a decreasing number of available microphones, outperforming them on both metrics for $M>15$ microphones. Note that PINN achieves its best $\operatorname{NMSE}$ with $M=50$: to reach the same score, PIBI requires only 30 microphones instead. This indicates that the proposed PIBI method is more data-efficient than traditional PINNs. As discussed in Sec.~\ref{sec:proposed_method}, whereas PINN relies on collocation points in the reconstruction region $\Omega$, PIBI only requires integration points on the boundary: this results in a reduction of the dimensionality of the problem, facilitating the convergence of the network and increasing its robustness to a lower number of microphones.

We finally fix the frequency to $\SI{390}{\Hz}$ and the number of microphones to $M=50$, and vary the number of collocation and integration points for PINN and PIBI, respectively. We report the results in Fig.~\ref{fig:nmse_plot}c and Fig.~\ref{fig:ncc_plot}c. The performance of PIDL is also reported for reference (horizontal line): as PIDL does not require collocation or integration points, its performance is constant in this experiment. Both PINN and PIBI outperform PIDL for all considered cases. Moreover, the figures show that, for each considered number of collocation (integration) points, PIBI outperforms PINN on both metrics. 
Once again, PIBI relies on points lying on the boundary of $\Omega$. Thus, for a given choice of $N_\text{coll}=N_\text{int}$, the density of integration points used in PIBI is higher than that of collocation points used in PINN, leading to a better performance. In other words, PIBI demands less integration points than PINN requires collocation points, resulting in a complexity gain.

These results indicate the potential of the proposed approach for sound field reconstruction. Imposing a physical constraint on the boundary of the target domain, rather than on points inside of it, helps to reduce the complexity of the problem and leads to improved performance and data-efficiency over PINNs. Moreover, the proposed approach has the advantage of yielding physically interpretable results that always represent solutions to the Kirchhoff-Helmholtz boundary integral equation.

\section{Conclusions}
\label{sec:conclusions}
In this paper, we proposed a sound field reconstruction method based on a boundary-informed neural network. By describing the sound field using the Kirchhoff-Helmholtz boundary integral equation, we adopted a multi-layer perceptron to retrieve the boundary pressure distribution, then employed to reconstruct the sound field at arbitrary positions within the analysis region. Simulation results indicate that the proposed approach outperforms both physics-informed dictionary learning and traditional PINNs on a broad frequency range and is more robust to scarcity of available measurements. Moreover, it demands fewer integration points compared to the collocation points required by PINNs, leading to a reduced problem complexity. Future work will focus on the analysis of regions containing sources (i.e., where the inhomogeneous Helmholtz equation holds) and on the introduction of explicit boundary conditions in the model. Finally, tests on real data will be carried out.

\bibliographystyle{ieeetr}
\bibliography{references}

\end{document}